# Mars surface phase function constrained by orbital observations

*Mathieu Vincendon*

*Institut d'Astrophysique Spatiale, Université Paris Sud, 91405 Orsay, France (mathieu.vincendon@ias.u-psud.fr, phone +33 1 69 85 86 35, fax +33 1 69 85 86 75)*

*Abstract:*

The bidirectional photometric properties of the surface of Mars describe how remote measurements of surface reflectance can be linked to hemispherical albedo used for energy balance calculations. A simple Lambert's law is frequently assumed for global data processing, even though several local studies have revealed the complexity of Mars surface phase functions. In this paper, we derive a mean Bidirectional Reflectance Distribution Function (BRDF) representative of widespread typical Martian terrains. OMEGA and CRISM orbital observations are used to provide observational constraints at solar wavelengths over a wide range of viewing conditions all over the planet. Atmospheric contribution is quantified and removed using a radiative transfer model. We observe a common phase behavior consisting of a 5 to 10% backscattering peak and, outside the backscattering region, a 10 to 20% reflectance increase with emergence angles. Consequently, nadir measurements of surface reflectance typically underestimate hemispherical reflectance, or albedo, by 10%. We provide a parameterization of our mean Mars surface phase function based on Hapke formalism ($\omega$ = 0.85, $\theta$ = 17, c = 0.6, b = 0.12, $B_0$ = 1 and h = 0.05), and quantify the impact of the diffuse illumination conditions which change surface albedo as a function of local time and season. Our average phase function can be used as a refinement compared to the Lambertian surface model in global data processing and climate modeling.

## 1 Introduction

The surface phase function characterizes the distribution of scattered radiation intensity with incidence and emergence directions. It is a function of surface texture (grain size, roughness, topography, depending on spatial scale), surface composition (complex refractive index), and illumination conditions (incidence angle, or distribution of incidence angles for a diffuse source). The surface phase function links single-direction reflectance to hemispherical reflectance integrated over all emergence directions. While most remote measurements of Mars surface reflectance are performed in a single viewing geometry (usually nadir), hemispherical reflectance is the physical quantity that measures the balance between incoming solar flux and scattered power in all directions, a critical value for climate modeling. For a perfectly Lambertian surface, the reflectance factor measured for any viewing geometry is equal to the hemispherical reflectance ("albedo"), a simplifying assumptions frequently used on Mars.

Estimating the surface phase function at Mars is a thorny issue. On the one hand, remote multi-angle measurements are sensitive to both surface and atmosphere, the latter being frequently dominant (Clancy & Lee 1991; Clancy et al. 2003; Vincendon et al. 2008; Wolff et al. 2009; Fernando et al. 2012; Shaw et al. 2012). Hence, complex modeling strategies are required to disentangle atmosphere and surface (Ceamanos et al. 2012) as estimations performed without removing the atmospheric contribution (Thorpe 1977; Pleskot and Kieffer, 1977; de Grenier & Pinet 1995; Soderblom et al. 2006; Jehl et al. 2008) can result in ambiguous functions. On the other hand, surface measurements performed by Landers (Guinness et al. 1997; Johnson et al. 1999; Johnson et al. 2006a,b) are restricted to a limited number of sites and are not necessarily representative of remote measurements of km-sized pixels which average various surface materials, textures and local topographies.

In this paper, we use observations obtained for various viewing geometries by the visible and near-IR imaging spectrometers OMEGA and CRISM to constrain over solar wavelengths the average phase function representative of typical Martian terrains observed at low spatial resolution (pixels > km). We account for the contribution of the atmosphere to both derive surface-only phase functions and study the impact of the changing diffuse illumination conditions through the spread of incidence directions over which the BRDF is integrated. Using Hapke formalism, we parameterize a mean Mars surface phase function consistent with these new data and the range of previously published constraints. This function provides a second order, better than the first order Lambertian surface model, surface phase function that can be used to integrate global data and compute albedo for surface energy balance calculation.

## 2  Method

### 2.1  Characterization of the surface phase function

We briefly define in this paragraph the physical quantities subsequently used, following conventions of Hapke (1993). The surface phase function refers to how a given surface scatters light as a function of emergence and incidence directions. The incidence direction is defined by the incidence angle *i*, or solar zenith angle (equal to 0° when the Sun is at zenith and 90° at the terminator). The emergence direction is defined by the emergence angle *e*, ranging from 0° toward zenith (i.e., for "nadir viewing") to 90°, and the azimuth angle. The phase angle *g* is the angle measured from the surface between emergence and incidence directions, ranging from 0° to 180°; it can be calculated from the three previous angles (see equation 8.4 of Hapke (1993)). The phase function is described by the Bidirectional Reflectance Distribution Function, *BRDF(i,e,g)* (see e.g. paragraph 10.B of Hapke (1993)). The BRDF is the ratio of the radiance scattered by a surface into a given direction to the collimated power incident on a unit area of the surface. Integrating the BRDF over all emergence directions provides the hemispherical reflectance (see e.g. paragraph 10.D.2 of Hapke (1993)), which corresponds to the total power scattered into the upper hemisphere by a unit area over the collimated power incident on this unit area. The hemispherical reflectance thus corresponds to the fraction of incoming energy not absorbed by the surface ("albedo"); it depends on solar zenith angle. For a Lambertian surface, the BRDF is constant, and the hemispherical reflectance is simply equal to *π x BRDF* and does not depend on incidence angle. By analogy, the reflectance factor REFF(i,e,g) = π x BRDF(i,e,g) is frequently used to characterize the reflectance in one direction (see e.g. equation 10.3 of Hapke (1993)).

We will use the same Hapke formalism as used in recent studies (Johnson et al. 2006a, b; Wolff et al. 2009; Fernando et al. 2012) to provide a parameterization of the phase function: the BRDF corresponds to the reflectance of equation 12.55 of Hapke (1993) divided by cos(i). Parameters are: opposition effect magnitude ($B_0$) and width (h), macroscopic roughness (θ), single scattering albedo (ω), and the asymmetry parameter (b) and backward fraction (c) of the 2-terms Henyey-Greenstein (HG) function. We use the same formalism as Johnson et al. (2006a) for c, i.e. c = (1+$c_{hapke}$) / 2 with $c_{hapke}$ corresponding to the "c" of equation 6.18a of (Hapke 1993).

## 2.2 OMEGA and CRISM data

The OMEGA (*Observatoire pour la Minéralogie, l'Eau, les Glaces, et l'Activité*) onboard Mars Express and CRISM (*Compact Reconnaissance Imaging Spectrometers for Mars*) onboard Mars Reconnaissance Orbiter instruments are imaging spectrometers observing the sunlight reflected by the surface and atmosphere of Mars at visible and near-IR wavelengths for various viewing geometries (Bibring et al. 2005; Murchie et al. 2007). Neither instrument performs complete measurements of the BRDF of the surface; nevertheless, their specificities make it possible to probe complementary portions of the surface phase function.

OMEGA mostly observes the surface of Mars with a near-nadir pointing direction and various incidence angles ranging from zenith to terminator depending on local time, season and latitude. Most places on Mars have been observed several times with various incidence angles since the beginning of OMEGA operations in early 2004. We can thus probe the BRDF for *e* = 0° and *i* between 0° and 90° by constructing time series of OMEGA observations. The spatial resolution of OMEGA ranges from 0.3 to 5 km depending on spacecraft altitude along its elliptical orbit. Time series of overlapping observations are constructed by looking at all observations obtained within homogeneous areas of a few tens to a few hundreds of km. Off-nadir (± 20° emergence) observations occasionally obtained by OMEGA were not used in our study. A dozen of "spot pointing" observations of the surface with varying emergences have also been acquired by OMEGA; however, preliminary modeling results (Vincendon et al., 2007b) indicate that these observations are dominated by aerosols effects, which makes them poorly suited for our surface analysis.

CRISM targeted observations are obtained by pointing to a target area on the ground with emergence angles ranging from 70° to 0° to 70° as the spacecraft flies over the target. The local time of observations is 15.00. These "Emission Phase Function" (EPF) observations thus probe the BRDF for a constant mid-afternoon solar zenith angle, an emergence angle varying from 0° to 70° for two opposite azimuth angles. CRISM targeted observations cover an area about 10x10 km² wide on the ground with a spatial resolution decreasing with emergence from 20 m at nadir to approximately 450 meters at 70° (Ceamanos et al., 2012); we sum all pixels within the covered area and thus derive CRISM EPF with spatial extent comparable to OMEGA time series.

## 2.3 Atmospheric contribution

Observations of the surface of Mars by remote sensors are troubled by the scattering and absorption of light within atmospheric aerosols and gas. Gas influence can be easily bypassed by selecting wavelengths over which gas is transparent, which is the case for a major fraction of the visible and near-IR range. Both dust and ice particles compose Mars atmospheric aerosols. Ice forms localized clouds which can largely be isolated and removed in the dataset using climatology database (Smith 2004) and spectroscopic evidence contained in near-IR OMEGA and CRISM data itself (Langevin et al. 2007). On the contrary, dust aerosols are ubiquitous: even under "clear" atmospheric conditions and favorable illumination conditions, at least 30% of solar photons interact with dust aerosols, with substantial effect on observed surface reflectance (Lee & Clancy 1990).

We use the multiple scattering Monte-Carlo radiative transfer code, aerosols optical properties, and methodology of Vincendon et al. (2007a, 2009) to model the contribution of dust aerosols to OMEGA and CRISM data. Look-up tables of apparent reflectance as a function of surface reflectance, and conversely, are produced as a function of viewing geometries and aerosols optical depths. The resulting atmospheric correction makes it possible to relate reflectance seen through the aerosols layer in a given direction to surface reflectance. The model assumes a Lambertian surface at the bottom boundary: the impact of this simplifying hypothesis will be discussed with other aerosols assumptions in section 3.3.

We also need to account for the non-collimated nature of incoming radiations due to aerosols scattering: as the BRDF depends on incidence angle, the diffuse Martian sky modifies the distribution of light scattered by the surface compared to a single incidence angle. The distribution of incoming illumination angles, as results from the dispersal of a single incoming solar zenith angle, is simulated using the same atmospheric radiative transfer code (see e.g. figure 5 of Vincendon et al. (2009)). Optical depths at time of observations are derived from Mars Exploration Rover (MER) measurements (Lemmon et al. 2004) performed during the Mars Express and Mars Reconnaissance Orbiter missions. The corresponding BRDF is then built by averaging the BRDF according to the modeled distribution of incidence angles.

## 2.4 Surface phase properties of individual samples

The phase functions of various Mars surface materials have been measured at the MER landing sites, and parameterized using Hapke formalism (Johnson et al. 2006a, b). Laboratory goniometric measurements on Martian analogues are also available (Shepard & Ward 2006; Souchon et al., 2011; Pommerol et al., 2011). We use these measurements to get a first overview of expected surface photometric behavior on Mars. We calculate on Figure 1 the relation between nadir reflectance and hemispherical reflectance for a representative set of materials at near-IR wavelengths (1 µm for MER, 0.75 µm for analogs). Both quantities should be equal for any incidence angle for a Lambertian surface. We model the non-collimated nature of incoming solar radiations caused by aerosols scattering (diffuse illumination, see section 2.3), which results in a reduction of the amplitude of surface phase variations with incidence angle. While there is a relatively high diversity of phase behavior, as expected from the diversity of material compositions and textures observed on Mars, some similarities are observed: most phase functions diverge significantly from Lambert's law (up to ± 20%), with a backscattering peak and, outside the backscattering direction, an increase of reflectance at high emergence angles. As a consequence, the reflectance measured at nadir outside the backscattering peak is always lower by 10 to 20% than the corresponding hemispherical reflectance (Figure 1). These measurements also show that the hemispherical reflectance is a function of the incidence angle, which is not the case for a Lambertian surface for which the albedo does not depend on illumination conditions (Figure 1): all surface materials analyzed by the MERs reflect 10 to 25% less radiation at low incidence angles compared to high incidence angles.

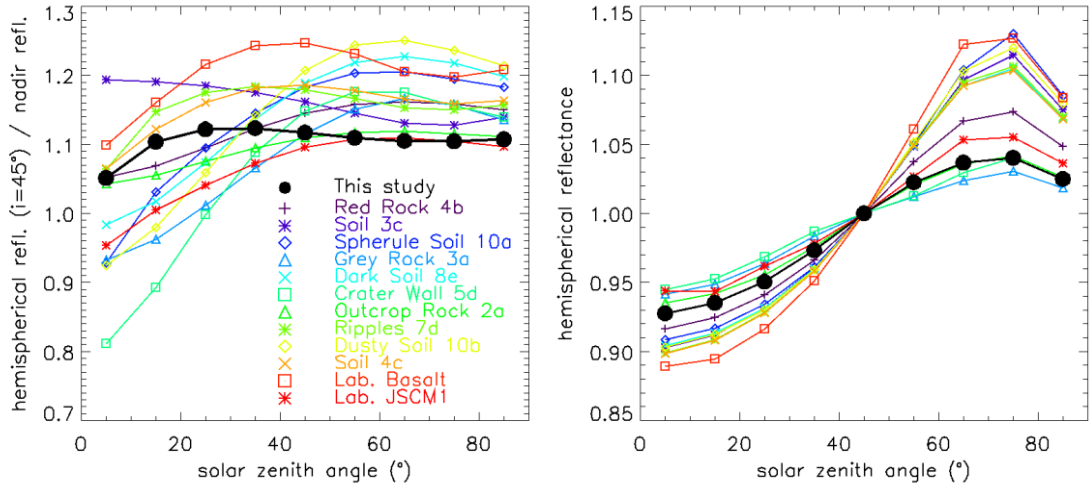

**Figure 1 : (Left)** Ratio between the hemispherical reflectance (at $i$ = 45°) and the observed nadir reflectance factor, as a function of incidence angle $i$. **(Right)** Hemispherical reflectance as a function of $i$, normalized to $i$ = 45°. The mean surface phase function derived in this study (black dots) is compared to phase functions observed at MER landing sites (colors, $\lambda$ = 1 µm, Johnson et al., 2006a, b) and to two laboratory measurements on analogs (red, 0.75 µm, Pommerol et al., 2012). Shown relations account for the diffuse illumination caused by an aerosols optical depth of 0.7. The Lambertian surface model would correspond to unity horizontal lines on these graphs. These measurements show that a wide diversity of local surface phase behavior is expected on Mars, with some relatively common trends: a backscattering peak and a nadir reflectance that underestimates the hemispherical reflectance. The average phase function derived in this study (black dots) corresponds to low-amplitude phase effects compared to most of these samples, as expected from the mixing of various textures, materials and local topographies within large areas.

MER measurements have been obtained for several visible wavelengths (Johnson et al. 2006a, b). In Figure 2, we show the same samples as in Figure 1, but for a wavelength of 0.43 µm located in the bottom of the ferric oxides absorption feature: the reflectance factor at 0.43 µm is about 0.05, significantly lower compared to the bright continuum reflectance factors of 0.15 – 0.4 measured at 1 µm. Significant variations of the phase behavior can be expected from theory when the absorption increases, such as an increase in relative terms of the backscattering peak intensity (Hapke, 1993). Overall, we can see that the spread of phase behavior for the various surface types considered is relatively similar at both wavelengths, with perhaps a slight average increase of the relative amplitude of phase variations at low reflectance levels ($\lambda$ = 0.43 µm, Figure 2) compared to high reflectance levels ($\lambda$ = 1 µm, Figure 1). For a given surface, we observe either a relatively constant phase behavior with wavelength (e.g., "Crater Wall 5d" with only a slight increase of the backscattering peak from 31% at 1 µm to 36% at 0.43 µm), significant relative increase of the intensity of the backscattering peak (e.g., "Soil 4c" from 10% at 1 µm to 30% at 0.43µm) or on the contrary significant decrease of the intensity of the backscattering peak (e.g., "Dusty Soil 10b", from 26% at 1 µm to 10% at 0.43 µm). Laboratory measurements on analogues perform by Pommerol et al. (2011) generally show a moderate increase of the relative intensity of the backscattering peak with decreasing wavelength (e.g., "Basalt", from 22% at 0.75 µm to 32% at 0.45 µm).

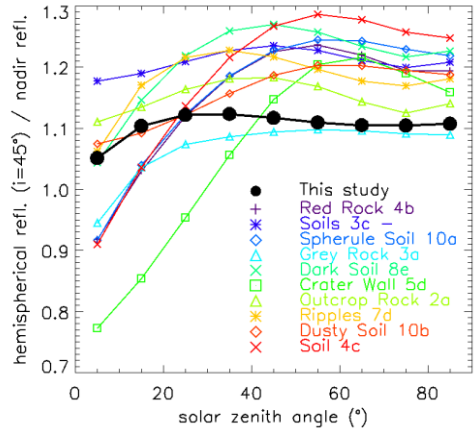

Figure 2 : Same as Figure 1 (left panel) for MER measurements at 0.43 µm, in the bottom of the surface ferric oxides absorption feature. Overall, no strong differences in the shape and strength of phase effects are observed when comparing low reflectances (0.43 µm, this figure) versus bright continuum reflectances (1 µm, figure 1). When looking at surfaces individually, we observe either weak changes (e.g., "Crater Wall") or significant increase or on the contrary decrease (e.g., "Soil 4c"/ "Dusty Soil 10b") of the relative intensity of the backscattering peak.

# 3   Results

## 3.1   OMEGA observations

We first analyze OMEGA aerosols-corrected series of nadir observations obtained with variable solar zenith angles. As overlapping observations have been obtained over more than 3 Martian Years (MY), we focus on places where minor surface changes have been previously reported (Szwast et al. 2006). We also focus on bright regions, for which the impact of aerosols on observed radiance is significantly lower (Clancy & Lee 1991). Observations contained in 8 homogeneous areas of a few degrees square are plotted versus solar zenith angles in Figure 3. The properties of these 8 areas are summarized in Table I. Most of these regions are located at low latitudes to get constraints on low solar zenith angles. Reflectance factors at 1.1 µm are shown. These observations reveal a relatively common surface phase behavior for all sites (Figure 4): the reflectance factor is 5 to 10% higher at 0 incidence (backscattering peak) compared to 25°, very slightly increases between 25° and ~ 60° (+ 2%) and is then relatively constant up to 80° considering the uncertainty on aerosols correction at high solar zenith angles. These observations show that low-level phase effects characterize the surface of Mars observed at nadir with various incidence angles: ±5% compared to the constant reflectance factor of a Lambertian surface.

We have failed to detect such phase effects for dark regions: due to the increased relative contribution of aerosols over low reflectance surfaces, the resulting incidence angle variations are significantly noisier compared to bright regions. Surface phase effects for dark regions, if present, must be of low amplitude so as to be masked by the noise resulting from aerosols uncertainties. This is consistent with conclusions of Vincendon et al. (2009) who demonstrated that seasonal and photometric variations over dark regions are primarily due to aerosols. Typically, similar low amplitude phase effects as for bright regions (< 10%) are below our aerosols removal uncertainty range and would be undetectable, while stronger phase effects would become apparent. Thus, while looking at the nadir reflectance as a function of incidence, we can conclude that dark regions are probably characterized by phase effects of similar (or weaker) amplitude as bright regions.

Table 1 : Properties of the eight homogenous areas used to construct time series of OMEGA nadir observations with various incidence angles (Figure 3). The number of usable OMEGA observations within each area and the corresponding latitude, longitude, incidence, $L_s$ and MY ranges covered are indicated. Observations are generally regularly distributed over the indicated $L_s$ range. All available MY and Ls are used, except for (e): MY27, $L_s$ < 270° only (the surface has changed afterwards). Offset and multiplicative factors used in Figures 3 and 4 are summarized in the last two columns.

| # | Name | Longitude range (°E) | Latitude range (°N) | Incidence range (°) | Number of observation used | Ls range covered (°) | Mars Year used | Offset of Figure 3 | Multiplicative factor of Fig. 4 |
|---|------|---------------------|---------------------|--------------------|-----------------------------|---------------------|----------------|--------------------|----------------------------------|
| a | Amazonis | 207 – 209 | 12 – 21 | 2 – 75 | 16 | 15 – 315 | 27 to 29 | -0.08 | 0.950 |
| b | Arabia | 22 – 26 | 14 – 18 | 9 – 81 | 25 | 5 – 285 | 27 to 30 | -0.06 | 1.000 |
| c | Elysium | 165 – 170 | -4 – 0 | 14 – 83 | 32 | 0 – 360 | 27 to 30 | -0.05 | 1.035 |
| d | Meridiani | 353 – 354 | 2 – 3 | 15 – 71 | 21 | 35 – 350 | 27 to 29 | 0.08 | 1.150 |
| e | Lunae | 290 – 299 | 8 – 11 | 5 – 78 | 17 | 10 – 265 | 27 | 0.08 | 1.130 |
| f | Isidis | 96 – 105 | 10 – 17 | 9 – 84 | 34 | 10 – 360 | 26 to 30 | 0.08 | 1.120 |
| g | Utopia | 150 – 153 | 42 – 44 | 28 – 75 | 13 | 0 – 360 | 27 to 30 | -0.05 | 1.017 |
| h | Daedalia | 207 – 212 | -15 – -11 | 2 – 83 | 16 | 15 – 340 | 26 to 28 | -0.06 | 0.970 |

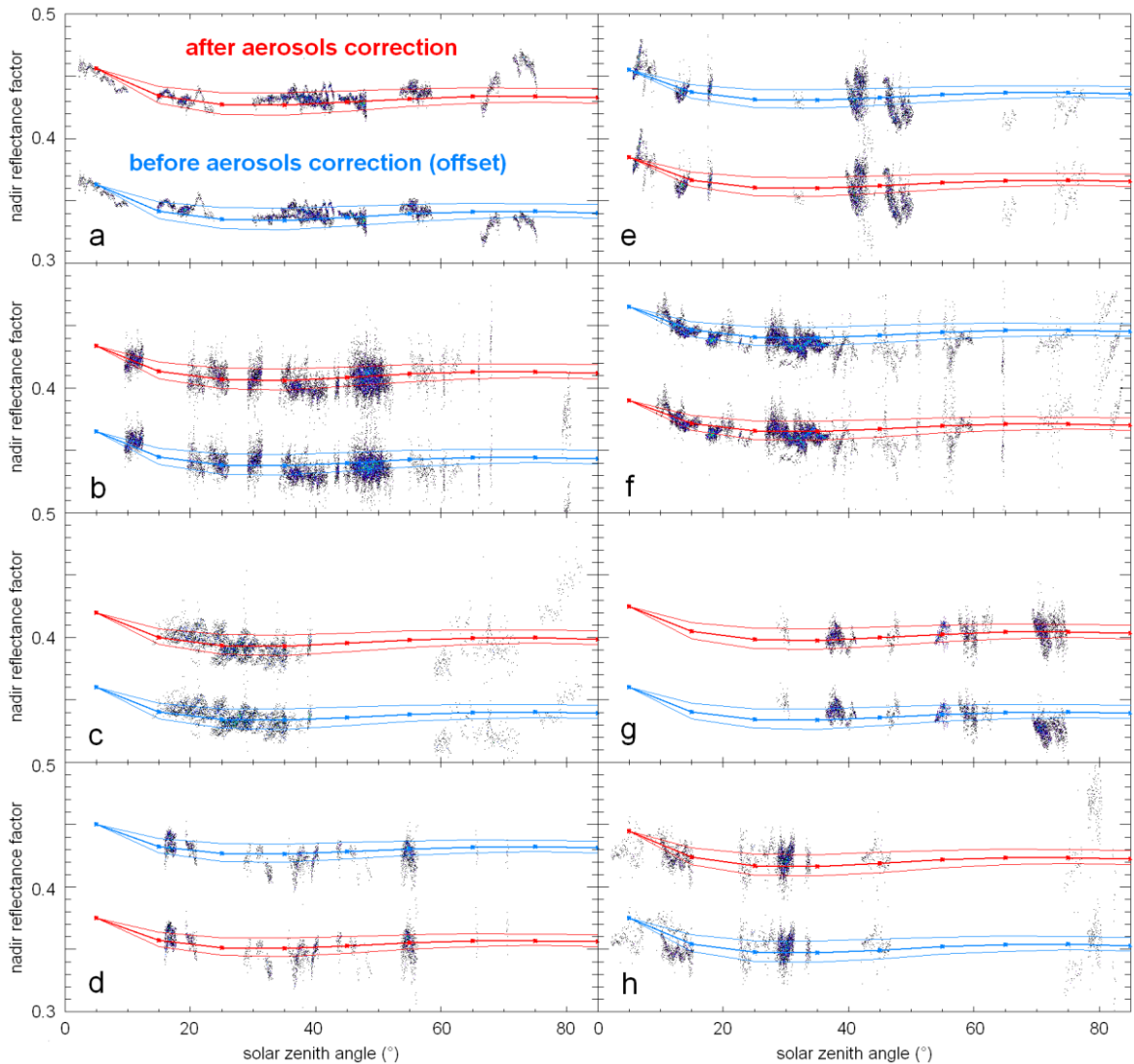

Figure 3 : OMEGA nadir surface reflectance factor at 1.1 µm as a function of solar zenith angle for eight bright areas stable over Martian Years (see areas description in Table 1). Data are shown after aerosols removal and compared to the average surface phase function derived in this study scaled with the appropriate multiplicative factor (red lines). Three aerosols optical depths τ covering the range of diffuse illumination conditions over which the data have been collected are shown: τ = 0.7 (red thick line) and τ = 0.3 and 1.4 (red thin lines). This makes it possible to appreciate the surface phase curve variability as a function of optical depth, to be compared to the observed variability among observations obtained under various aerosols optical depth. Observations prior to aerosols removal are also shown with an offset (phase function in blue, offsets indicated in Table 1) to assess the impact of the aerosols correction.

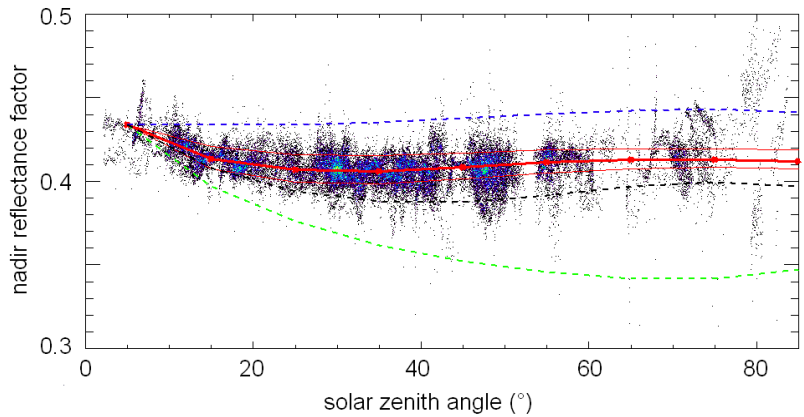

Figure 4: OMEGA surface nadir reflectance factor as a function of solar zenith angle: mix of the eight areas of Figure 3, with appropriate multiplicative factors (Table 1). Red: surface phase function derived in this study (thick line: τ = 0.7; thin lines: τ = 0.3 / 1.4). Observations are compared to previous estimates of Mars surface phase function (dashed lines, normalized to *i* = 5°). The amplitude and width of the Wolff et al. (2009) backscattering peak (green) are three times greater than observed. Functions typical of Fernando et al. (2012) retrievals (blue) are consistent with our observations for *i* > 20° but not appropriate at lower incidence as they do not constrain the opposition effect. The closest MER function (right shape but slightly higher amplitude) corresponding to Spirit soil 6c (Johnson et al. 2006a) is indicated in black.

These surface phase effects are generally constant over visible and near-IR wavelengths. For example, the intensity of the backscattering peak of area "f" (Figure 5) is constant in % for wavelengths corresponding to both high reflectance levels (continuum, near-IR) and low reflectance levels (ferric oxides absorption band, short visible wavelengths). This result is in agreement with the overall stability of phase effects with wavelengths seen in MER measurements (section 2.4), as we probably average several surface materials within our large studied areas. For area "a", we observe an increase in the relative intensity of the backscattering peak with decreasing wavelength, as seen in some individual MER and laboratory analog samples (see section 2.4). However, it is difficult to disentangle surface and atmospheric effects at early visible wavelengths due to the complex spectral behavior of dust aerosols, the possible contribution of small-grained water ice and the increasing contribution of Rayleigh scattering (Wolff et al. 2009, 2010).

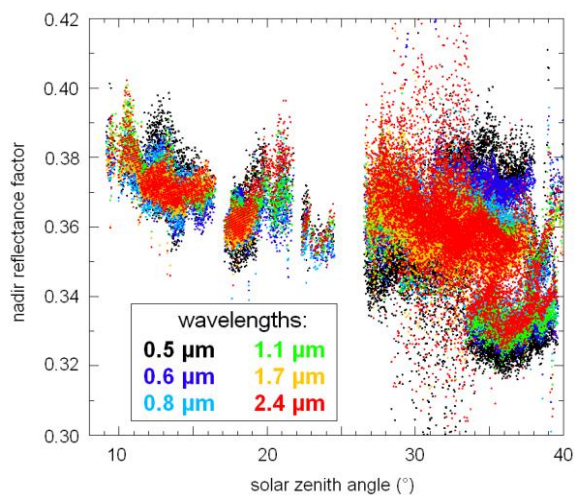

Figure 5 : Phase curve of area f is compared for different visible and near-IR wavelengths. The reflectance spectrum of Mars is characterized by a broad ferric absorption with reflectances four times lower at short visible wavelengths compared to near-IR wavelengths. As a consequence, multiplicative scaling factors are applied to compare wavelengths: 3.7 (0.5 μm), 1.59 (0.6 μm), 1.08 (0.8 μm), 1.01 (1.1 μm), 0.985 (1.7 μm) and 1.0 (2.4 μm). The relative amplitude of the surface backscattering peak is similar, in %, for all wavelengths.

## 3.2 CRISM observations

We analyze a few CRISM Emission Phase Function ("EPF") observations and integrate results from ongoing CRISM EPF modeling by other teams (Wolff et al. 2009; Fernando et al. 2012). As representative examples, we have selected two CRISM observations obtained under relatively clear atmospheric conditions (visible optical depth between 0.4 and 0.6) and corresponding to two very distinct surface types: EPF # FRT9901 over a bright region with local topography near Valles Marineris (Figure 6), and EPF # FRT50B2 over a very homogeneous and flat dark region at Syrtis Major (Figure 7). CRISM EPF observations show that the "surface + atmosphere" reflectance increases with emergence angle, compared to a flat Lambertian surface. At 1.1 µm, observed reflectance variations with emergence (0-70° range) and phase (40-100° range) angles are ± 10% for the bright region and ±50% for the dark region. To help disentangling surface and aerosols effects, we explore CRISM phase curves at 1.1 and 2.4 µm, as the contribution of aerosols decreases with increasing near-IR wavelength (Vincendon et al. 2009). We can see in Figure 6 that the phase curve of the bright region is dominated by surface effects and relatively constant between 1.1 µm and 2.4 µm, while aerosols effects dominate for the dark region, in particular at 1.1 µm. These observations show that the surface reflectance factor observed at nadir is lower than reflectance factors observed at non-zero emergence angle, which means that nadir reflectance factor is necessarily lower than hemispherical reflectance, similarly to what is observed by the MERs (section 2.4). The uncertainties related to the dominant contribution of aerosols over dark regions make it impossible to properly constrain a unique surface phase behavior there; however, we can observe that the phase function of bright regions is also appropriate for dark regions (Figure 7).

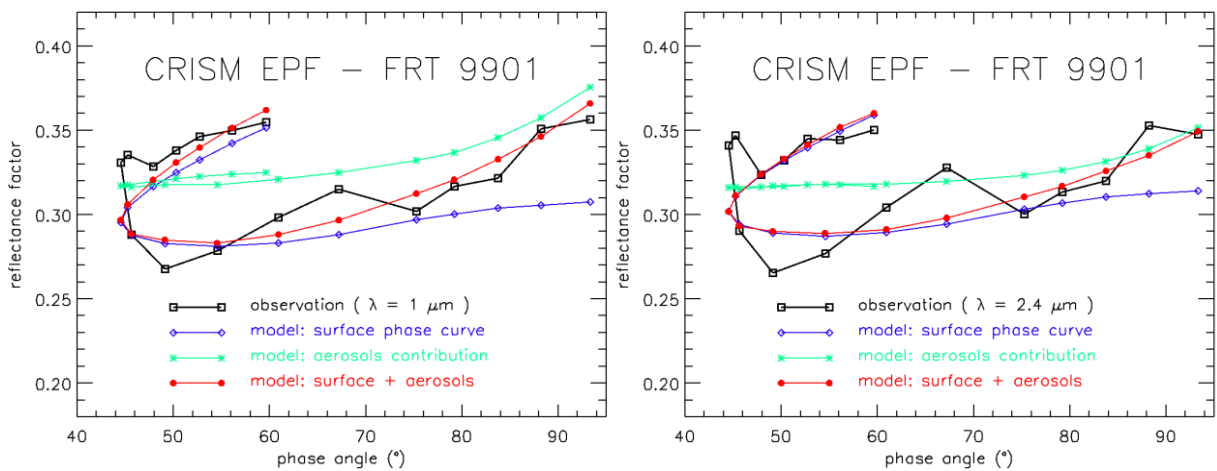

Figure 6 : CRISM Emission Phase Function (EPF) observation # FRT 9901 (bright region, 291°E, 6°S, Ls 20°, MY29). The solar zenith angle is 41°; the emergence angle ranges from -70° to +70°. Wavelengths 1 µm (left) and 2.4 µm (right) are shown. Black squares: observed reflectance factor as a function of phase angle. Blue diamonds: corresponding variations of the surface photometry using the phase function derived in this study. Green stars: corresponding aerosols contribution (optical depth of 0.5 at 1 µm). Red dots: resulting surface + aerosols phase curve. The contribution of aerosols is significantly lower at 2.4 µm (optical depth reduced by a factor of 2). The observed phase curve mainly results from surface phase effects, which are observed to be nearly identical at 1.1 µm and 2.4 µm. Slight changes in the observed area during the acquisition of the targeted observation may contribute to the observed distortion of the phase curve.

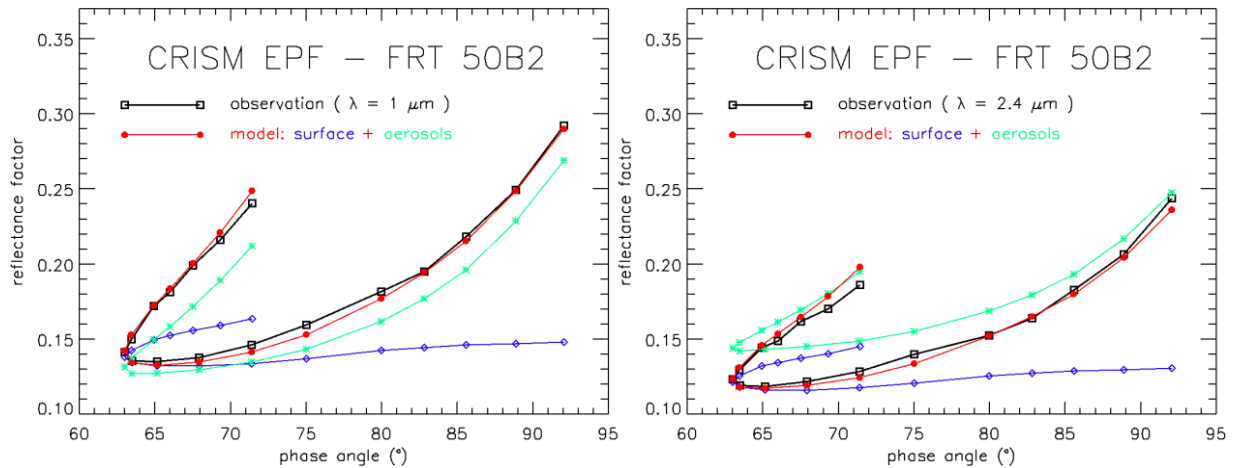

**Figure 7 :** Same as Figure 6 for CRISM observation # FRT 50B2 (dark region, 68°E, 9°N, Ls 209°, MY28). The solar zenith angle is 56°. Despite similar optical depth (0.55 at 1 µm) and viewing conditions compared to FRT 9901, the contribution of aerosols is significantly larger due to the darker underlying surface. The surface phase function derived for bright regions is also appropriate for this dark region; however, we must notice that other surface phase functions could also be appropriate considering the large aerosols contribution and associated uncertainties.

### 3.3 Impact of aerosols assumptions

We discuss here the impact of aerosols radiative transfer modeling assumptions on retrieved surface properties. Aerosols properties at a given wavelength are described by three parameters: optical depth, single scattering albedo, and single scattering phase function. The single scattering albedo is relatively well-constrained in particular for wavelengths > 0.65 µm (Vincendon et al., 2007; Wolff et al. 2009) and does not significantly impact our results. The assumed phase function is that of Tomasko et al. (1999), in agreement with results of Wolff et al. (2009). We assess the impact of changing this phase function by using a Henyey-Greenstein function with an asymmetry parameter of 0.63 (Ockert-Bell et al., 1997; Vincendon et al., 2007). We use optical depths corresponding to the average MER trends (see section 2.3), while a given observation can correspond to specific localized dust events. A change in aerosols optical depth by ± 50% is thus tested. The aerosols contribution for these various assumptions is shown in Figure 8 for CRISM EPF # 9901. Interpretation is poorly dependent on aerosols parameters: the observed phase variations at low to moderate phase angles remain mainly due to surface effects, while most of the increase of reflectance at higher phase angles is still due to aerosols.

We appreciate the impact of the aerosols correction on OMEGA data in a more drastic manner by comparing corrected data to data without any aerosols removal (Figure 3). At low solar zenith angles (*i* < 40°), the impact of the aerosols correction on the shape and intensity of the backscattering peak is barely discernible considering the spread in data points. The impact of the contribution of aerosols is important at higher incidence angle, where no significant variations from a constant line are observed if we consider both the spread observed in data points and the spread resulting from considering both corrected and un-corrected data. We use an aerosols phase function without backscattering peak, as constrained by observations (Tomasko et al., 1999). This explains why at low solar zenith angle, where the changing path length with incidence is moderate, aerosols do not significantly impact the phase curve. One could argue that theoretical models of geometric particles, such as Mie or T-matrix models, sometimes predict a backscattering peak; however, it has been shown recently that such theoretical backscattering peak has to be removed to match observations (Wolff et al., 2010).

Finally, our radiative transfer code assumes a Lambertian surface to relate reflectance with and without aerosols, which may appear not self consistent as we derive a non-Lambertian surface phase function (Ceamanos et al. 2012). It has notably been shown that assuming a Lambertian surface while doing the atmospheric radiative transfer does bias the estimation of surface photometric properties (Lyapustin, 1999; Fernando et al., 2012). The radiance measured in a given direction through the aerosols layer corresponds to various components: surface scattering without interaction with aerosols, single and multiple scattering by aerosols without interaction with the surface, and photons scattered by aerosols after being scattered by the surface. These latter photons will be impacted by the assumption on the surface phase function: photons leaving the surface with high emergence angles have e.g. a stronger probability to interact with aerosols and see their trajectory modified compared to photons leaving toward zenith. We used our Monte-Carlo model (Vincendon et al. 2007) to follow the path of photons through the atmosphere under typical conditions (surface albedo of 0.3, aerosols optical depth of 0.5, various incidence angles). Between 15% and 35% of observable photons are scattered by aerosols after surface scattering and thus sensitive to the surface phase function. As the aerosols phase function is strongly forward scattering, about half of these photons do not deviate by much than ±15° of their original direction. Thus, a maximum of 5% to 15% of collected photons in a given direction can originate from a significantly different direction. Observed and expected surface phase function typically results in ±10% variations in the amount of emitted photons with phase angles, which means that overall, we estimate that this effect corresponds to reflectance variations of typically 1-2 % maximum. This uncertainty is about one order of magnitude lower than observed surface phase variations and is also lower than typical uncertainties linked with the aerosols correction, as discussed previously.

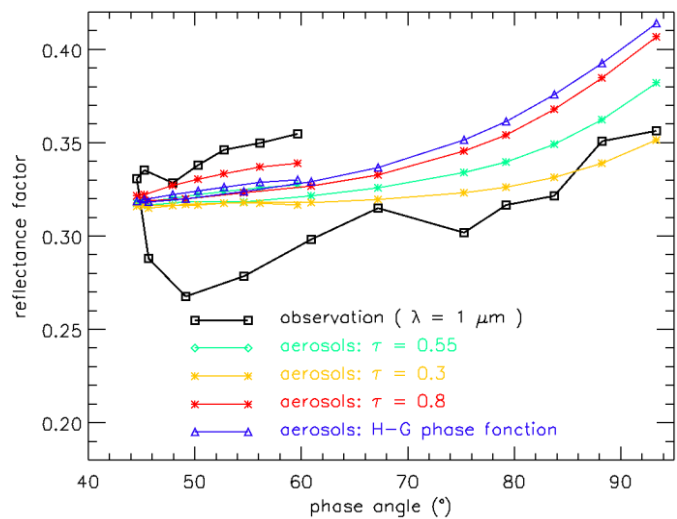

**Figure 8 : Impact of aerosols properties on the modeling of the aerosols contribution to the « surface-dominated » CRISM EPF 9901. Observation (black square, same as Figure 6 left) is compared to the aerosols contribution simulated for various hypotheses. (green stars) Tomasko et al. (1999) phase function with τ = 0.55 (nominal model, same as Figure 6 left). (yellow and red stars) Optical depth changed by ± 50% (0.3 and 0.8 respectively). (blue triangles) Henyey-Greenstein phase function (asymmetry 0.63) with nominal τ = 0.55. Results are poorly impacted by the choice on aerosols parameters: the contribution of aerosols remains low for phase angles < 70° used to constrain surface properties, while the contribution of aerosols to the reflectance increase at high phase angle remains large.**

## 3.4 Comparison with previous studies

We have compared our observational constraints to phase functions previously derived using CRISM and MER measurements. The mean CRISM phase function derived by (Wolff et al. 2009) deviates too much from Lambert's model compared to our observational constraints (Figure 4). The Wolff et al. (2009) surface phase function could be contaminated by a remaining aerosols component (see e.g. figure 1 of their paper), which would have no impact on the modeling of dust storm EPF performed in most of the paper, but which could explain why retrieved optical depths in their figures 18 and 19 are generally underestimated compared to MER measurements. Fernando et al. (2012) retrievals do not include the opposition effect due to a lack of constraints in CRISM EPF, which results in a relatively flat reflectance curve in the backscattering region (Figure 4). Finally, none of the local phase functions derived at MER landing sites exactly matches our observations (MER relations of Figures 1 and 2 can be simply compared to the OMEGA incidence curve of Figure 4, as they correspond to the same quantity, but reversed and multiplied by a constant factor).

## 3.5 Derivation of a mean surface phase function

In this paragraph, we present our approach to derive a mean surface phase function, parameterized with Hapke formalism (see paragraph 2.1), consistent with both our observational constraints and previous studies. We note that several sets of Hapke parameters can result in relatively similar phase functions consistent with our observations. However, we are not interested here in the physical interpretation of Hapke parameters.

The shape of the OMEGA phase curve is similar, but with lower amplitude, to a certain class of MER phase functions analyzed by Johnson et al. (2006a, b): Spirit Soils of table 6c and 4c, Opportunity Ripples of tables 3b and 7d, and Opportunity Dusty Soil of table 7c. These bright dusty surface types are indeed expected to be a major component of our large scale OMEGA bright regions. These surfaces share common values of HG-2 phase function parameters: relatively low $b$ of 0.1 to 0.2 (broad, low amplitude lobes) with high $c$ of 0.4 to 0.7 (backscattering surfaces). These surfaces are also characterized by an opposition effect with $B_0$ = 1 (its theoretical value) and with $h$ (lobe width) between 0.02 and 0.2. Similar $c$ values and slightly higher $b$ (0.2 – 0.3) have been found from previous CRISM EPF modeling (Wolff et al. 2009; Fernando et al. 2012). Starting from these ranges, we adjust the Hapke parameters to visually fit OMEGA and CRISM constraints of Figure 4, Figure 6, which results in the following set of parameters: c=0.6 (backscattering surface), b=0.12 (low amplitude lobes), $B_0$=1 and h=0.05 (narrow opposition effect). An increased macroscopic roughness θ in the 15 – 20° range is then required to decrease the amplitude of phase variations. Such a value is expected from the mixing of various surface types, textures and local topographies within large areas, and is consistent with values in the 15 – 18° range derived from CRISM EPF (Wolff et al., 2009; Fernando et al., 2012). A mean value of 17° is selected. Finally, we must set the single scattering albedo (ω) which controls the reflectance level of the surface: relatively high ω of 0.8 to 0.9 leads to our observed bright regions reflectance of 0.3 to 0.4 in the near-IR continuum. Decreasing ω results in an increase of the non-Lambertian behavior of the surface, similarly to decreasing the macroscopic roughness θ (ω=0.85 and θ=15° is e.g. relatively similar to ω=0.75 and θ=20°). Hence, a relatively high value of the single scattering albedo (ω = 0.8 – 0.9) is also required. A mean value of 0.85 is selected.

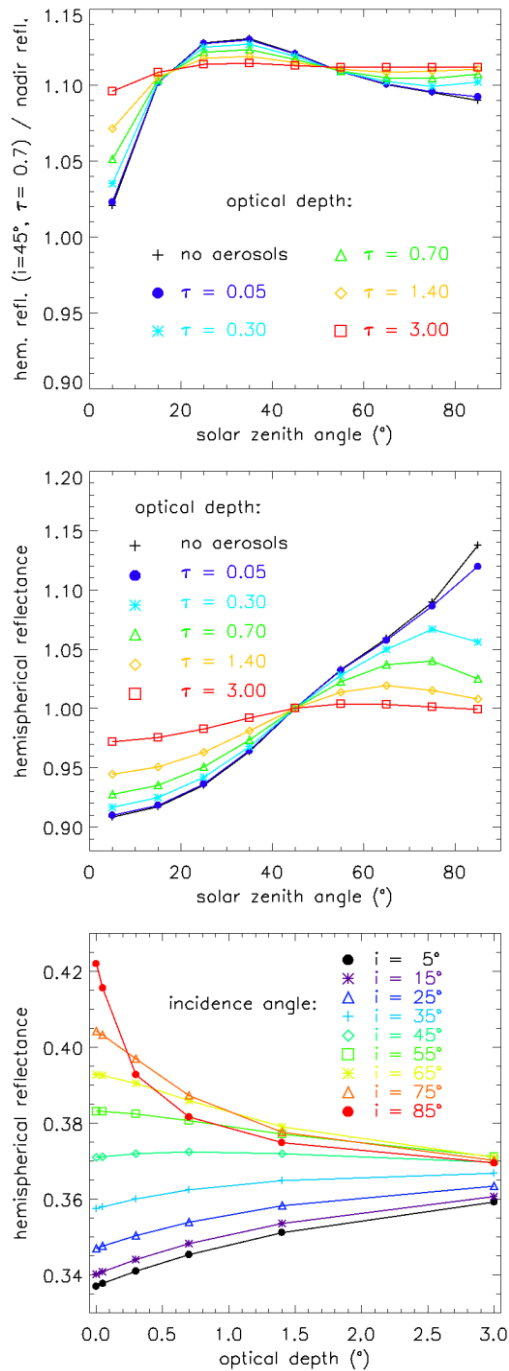

**Figure 9 : Useful relations derived from our mean surface phase function (ω=0.85, θ=17°, c=0.6, b=0.12, B0=1 and h=0.05), as a function of aerosols optical depth τ and incidence angle i. Top, relation between the nadir reflectance factor and the hemispherical reflectance at fixed reference average values of i and τ (i = 45°, τ = 0.7). Middle, hemispherical reflectance as a function of i for various τ, normalized to i = 45°. Bottom, hemispherical reflectance as a function of τ for various i. Increasing the aerosols optical depth reduces the collimated nature of incoming solar radiations, which decreases the variability of the surface phase function with solar zenith angle. Increasing the optical depth does however not impact the dissymmetry between nadir and higher emergences, the reason why the ratio between hemispherical reflectance and nadir reflectance is 1.1 at high optical depth, and not 1 (Lambert's law). The hemispherical reflectance at i=45° weakly depends on optical depth; i=45° is thus well suited when deriving global hemispherical reflectance maps.**

We present in Figure 9 useful relations derived from this phase function. Relations are shown for various illumination (i.e., aerosols optical depth) conditions, as the presence of aerosols in the atmosphere results in a spread of the distribution of incoming solar photons directions. The top panel shows how to relate nadir reflectance measurements to hemispherical reflectance, which makes it possible to link spacecraft reflectance measurements to hemispherical albedo needed for energy balance calculation. This first relation contains the backscattering peak at low solar zenith angle (< 20°); for larger incidence angles, the hemispherical reflectance is about 10% higher than the nadir reflectance factor. The bottom two panels show how evolves the hemispherical reflectance as a function of optical depth and solar zenith angle, which makes it possible to quantify the daily and seasonal variations of surface albedo in climate models. The hemispherical reflectance is higher at high incidence angles, as photons reaching the surface with a nearly horizontal direction have a lower probability to penetrate through the regolith. Increasing the aerosols optical depth increases the range of actual incidence angles for a given incidence at the top of the atmosphere, reducing the variability of the hemispherical reflectance with incidence. The hemispherical reflectance is nearly independent of the optical depth at the average incidence angle of 45°, while variations up to 15% are observed at extreme incidence angles; i=45° is thus an appropriate base value for a global albedo map of the surface of Mars. We have seen in previous sections that the relative amplitude of phase effects (at large spatial scale) is overall constant at first order over visible and near-IR wavelengths (e.g., Figures 5, 6) and for bright and dark terrains (e.g., Figure 7). Thus, we simply need to scale the relations of Figure 9 by a multiplicative factor to fit the average brightness of any considered terrain or wavelength.

# 4 Summary and conclusion

We used OMEGA and CRISM visible and near-IR data to constrain the mean surface phase function of Mars as averaged over large (> 10km) areas. Aerosols scattering is accounted for through the use of a radiative transfer code. We observe a relatively common average surface phase behavior independently of observed area and wavelength, composed of a slight backscattering peak and, outside the backscattering region, an increase of reflectance with emergence angles. This average behavior is consistent with both in-situ measurements at the surface of Mars and laboratory characterizations of Mars soil analogs. Some variability around this average behavior is observed, as expected from the diversity of Mars surface types and the variability of refractive indices over the solar range; however, it has not been possible to reliably quantify any particular trend as observed changes are within our atmospheric contribution uncertainties range. Polar icy surfaces were not included in our study: bidirectional reflectance functions of such surfaces may differ from ice-free terrains and may vary over seasons (Domingue et al. 1997; Bourgeois et al., 2006).

Using these observations and the range of constraints previously published, we have derived a mean phase function that represents a second order model compared to Lambert's law when processing global reflectance data of Mars. This phase function is parameterized using Hapke formalism (same conventions as Johnson et al. (2006a, b)) with the following set of parameters: $\omega = 0.85$, $\theta = 17$, $c = 0.6$, $b = 0.12$, $B_0=1$ and $h=0.05$. Relations derived from this phase function for various illumination conditions (various aerosols optical depths) are presented. Considering this phase function results in three main effects: (1) homogenize nadir measurements obtained at different solar zenith angles by accounting for a 10% backscattering peak at low solar zenith angles (< 20°); (2) account for the scattering increase with emergence angle by multiplying nadir reflectance by about 1.1 to get average hemispherical reflectance; and (3) account for a 10% increase of the hemispherical reflectance with solar zenith angle, and for the variability of the hemispherical reflectance with aerosols optical depth. These refinements makes it possible to account for the non-Lambertian properties of the surface of Mars while combining remote data obtained under various viewing geometries. Single direction reflectance measurements can then be linked to the hemispherical reflectance or albedo. Moreover, radiative budget calculation accounting for these surface phase effects can be integrated in general circulation models to enhance climate modeling precision, as surface albedo actually varies as a function of local time and season via its dependence on solar zenith angle and diffuse sky illumination conditions.

# 5   Acknowledgment

The author would like to thank the OMEGA and CRISM engineering and scientific teams for their help and for making this work possible. We also thank M. J. Wolff, A. Pommerol and C. Pilorget for interesting discussions, as well as two anonymous Reviewers for their meticulous reading of the manuscript.